\documentclass[conference]{IEEEtran}

\usepackage[english]{babel}
\usepackage[rflt]{floatflt}
\usepackage{graphicx}
\usepackage{url}
\usepackage{multirow}
\usepackage[rflt]{floatflt}
\usepackage{graphicx}
\usepackage{multirow}
\usepackage[table]{xcolor}
\usepackage{supertabular}
\usepackage{spverbatim}
\usepackage{multicol}
\usepackage{csquotes}
\usepackage{flushend}
\usepackage{cite}

\usepackage{enumitem}

\usepackage{todonotes}
\usepackage{trackchanges}

\newcommand{\x}{\times}

\addeditor{DS}

\title{Security and Performance Comparison of Different Secure Channel Protocols for Avionics Wireless Networks}

\begin{document}

\author{\IEEEauthorblockN{Raja Naeem Akram\IEEEauthorrefmark{2}, Konstantinos Markantonakis\IEEEauthorrefmark{2}, Keith Mayes\IEEEauthorrefmark{2}\\
Pierre-Fran{\c c}ois Bonnefoi\IEEEauthorrefmark{3}, Damien Sauveron\IEEEauthorrefmark{3}\IEEEauthorrefmark{4} and Serge Chaumette\IEEEauthorrefmark{4}
}
\IEEEauthorblockA{\IEEEauthorrefmark{2}Information Security Group Smart Card Centre, Royal Holloway, University of London, Egham, United Kingdom\\
\IEEEauthorrefmark{3}XLIM (UMR CNRS 7252 / Universit\'e de Limoges), D\'epartement Math\'ematiques Informatique. Limoges, France\\
\IEEEauthorrefmark{4}LaBRI (UMR CNRS 5800 / Universit\'e de Bordeaux), Talence, France \\
Email: \{r.n.akram, k.markantonakis, keith.mayes\}@rhul.ac.uk,\\ \{pierre-francois.bonnefoi, damien.sauveron\}@unilim.fr,  serge.chaumette@labri.fr}}

     \maketitle

\begin{abstract}
The notion of Integrated Modular Avionics (IMA) refers to inter-connected pieces of avionics equipment supported by a wired technology, with stringent reliability and safety requirements. If the inter-connecting wires are physically secured so that a malicious user cannot access them directly, then this enforces (at least partially) the security of the network. However, substituting the wired network with a wireless network - which in this context is referred to as an Avionics Wireless Network (AWN) - brings a number of new challenges related to assurance, reliability, and security. The AWN thus has to ensure that it provides at least the required security and safety levels  offered by the equivalent wired network. Providing a wired-equivalent security for a communication channel requires the setting up of a strong, secure (encrypted) channel between the entities that are connected to the AWN. In this paper, we propose three approaches to establish such a secure channel based on (i) pre-shared keys, (ii) trusted key distribution, and (iii) key-sharing protocols. For each of these approaches, we present at least two representative protocol variants. These protocols are then implemented as part of a demo AWN and they are then compared  based on performance measurements. Most importantly, we have evaluated these protocols based on security and operational requirements that we define in this paper for an AWN.
\end{abstract}

\section{Introduction}
In today's aircraft, Aircraft Data Networks (ADNs) -- highly reliable, efficient and fault-tolerant distributed (real-time) networks -- interconnect a large number of avionics sub-systems, enabling data and network management commands (control messages) to be exchanged within a predefined and deterministic time frame. These ADNs have to cater for a number of sub-systems with both critical and non-critical functions. Building a network that efficiently manages these two functions, while still providing a fully deterministic network with guaranteed bandwidth and Quality of Service (QoS), is extremely challenging \cite{Robinson2007a}. 

These ADNs are basically wired networks that connect multiple devices using a physical connection. Examples of such networks include ARINC 825 \cite{ARINC825}, ARINC 664/AFDX (Avionics Full DupleX Switched Ethernet) \cite{ARINC664,Safwat2014} and standard Ethernet. The wiring of these network cables requires an extensive design-time configuration of the aircraft, making post adaptation less flexible, not to mention the potential of wires being eroded and the problem of additional weight. In addition, the network redundancy is based on dissimilar paths, not dissimilar mediums of communication, even though the latter is known to be a better solution.

For these reasons, potentially, for non-critical functions of an aircraft a wired link between two communication points can be replaced with a wireless communication medium. 
Such a network is referred to as an Avionics Wireless Network (AWN). The potential for wireless communication in the AWN to be eavesdropped and/or modified is comparatively higher than for an ADN. For this reason, all communication between wireless nodes (pieces of equipment) in an AWN should be encrypted. To achieve such secure communication (via encryption), AWN nodes have to establish secure channels between each other, by running a secure channel protocol. In this paper, we have selected seven such secure channel protocols based on three different wireless communication deployment approaches. These selected protocols are then analysed for their suitability from performance and security point of view for AWN deployment. 

\subsection{Contribution}
In this paper, our main focus is on the security and performance analysis of different secure channel protocols for AWNs.
Our salient contributions are the following:
\begin{enumerate}
	\item selection of seven secure channel protocols based on three different setup approaches (pre-shared keys, trusted key distribution, and key sharing protocols);
	\item definition of criteria to compare these secure channel protocols along with the related security and performance analysis;
	\item implementation of these protocols in a AWN test-bed based on off-the-shelve hardware, so as to be able to make measurements.
\end{enumerate}

\subsection{Structure of the Paper}
Section~\ref{sec:Aircraft_Data_Networks} briefly presents the generic architecture of Aircraft Data Networks, which constitutes a rationale for considering the benefits of using Wireless Networks. Section~\ref{sec:Avionics_Wireless_Network} discusses different AWN formats and the need for secure channel protocols and proposes a case study in which the secure channel protocols presented in section~\ref{sec:Establishing_Secure_Channels} may take place.
Section~\ref{sec:AWN_Deployment_Scenarios} introduces the studied secure channel deployment scenarios.
Section~\ref{sec:Establishing_Secure_Channels} provides the results of the evaluation of several secure channel protocols in a AWN test-bed. Section~\ref{sec:Security_and_Performance_Analysis} compares them based on the obtained results according to the defined security and performance criteria.
In section~\ref{sec:Conclusion} we provide a summary of our experiments.

\section{Aircraft Data Networks}
\label{sec:Aircraft_Data_Networks}
In this section, we discuss the traditional deployment of a wired network (ADN) in an aircraft. We conclude with a short list of benefits that would result from replacing some non-critical functions of ADN with wireless technology. 

\subsection{Generic Architecture}
\label{sec:Generic_Architecture}
\begin{figure}[htbp]
	\centering
		\includegraphics[width=1.00\columnwidth]{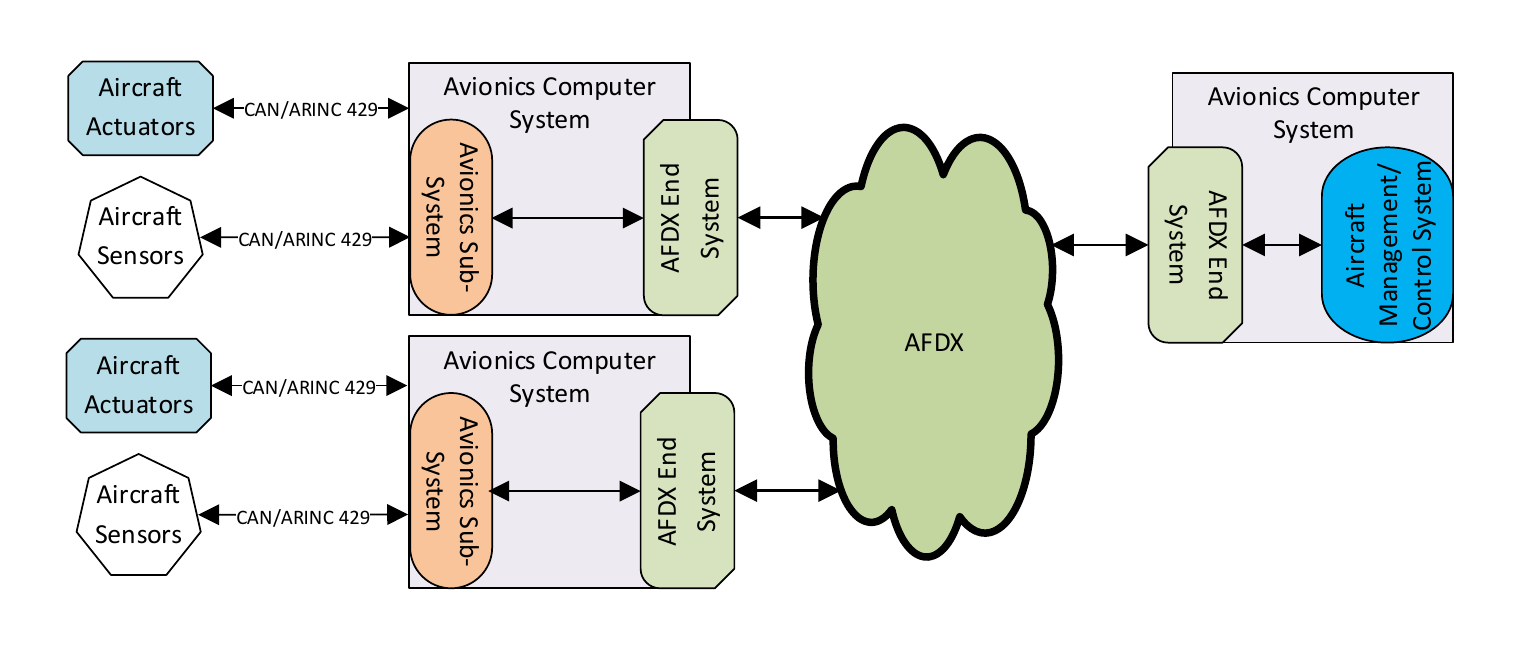}
	\caption{Generic Aircraft (Flight-Control and/or Crew) Data Network with AFDX as an Example}
	\label{fig:AircraftDataNetwork}
\end{figure}

\begin{figure}[htbp]
	\centering
		\includegraphics[width=1.00\columnwidth]{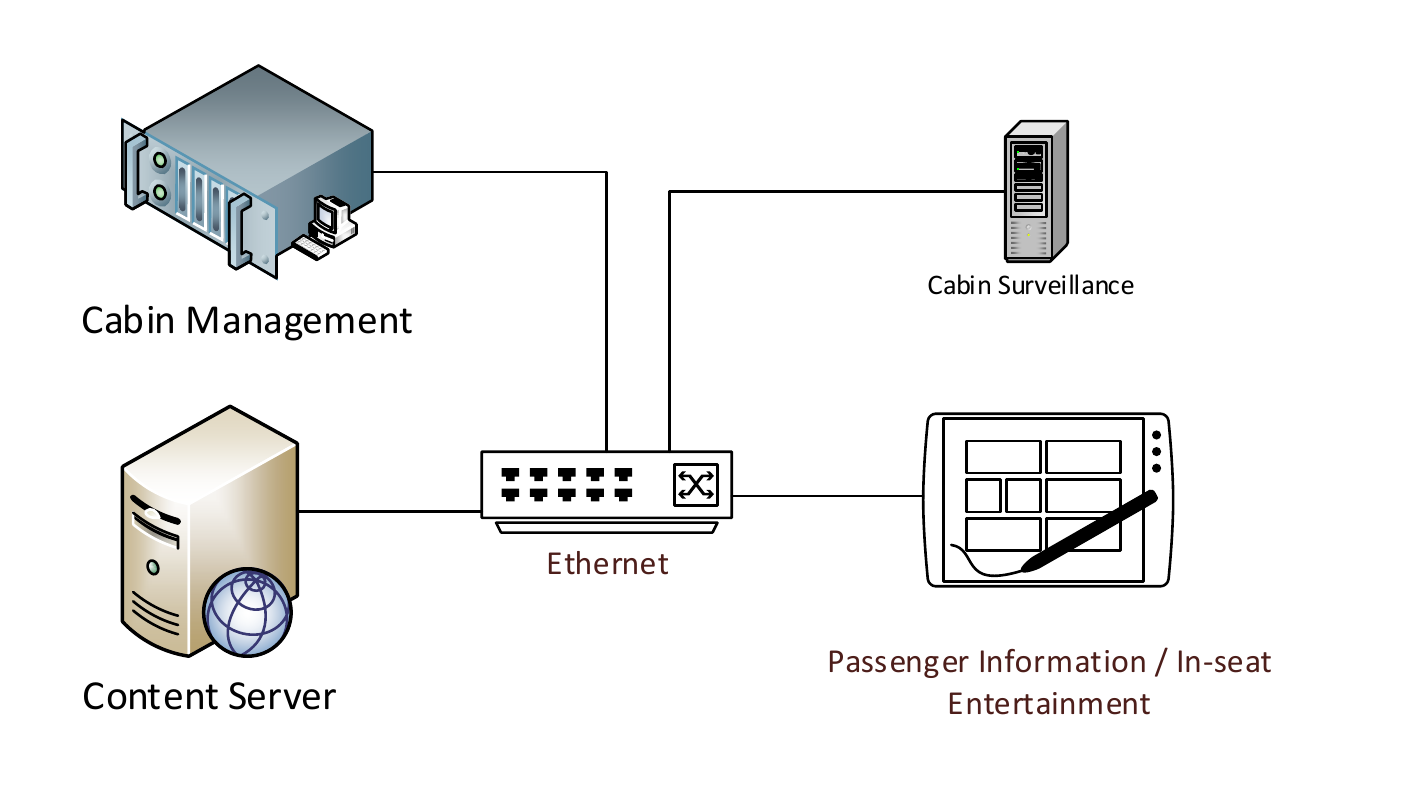}
	\caption{Generic Aircraft (Cabin) Data Network}
	\label{fig:Aircraft_Entertainment_Data_Network}
\end{figure}

A modern aircraft network consists of several data networks, including flight-control and/or
crew network (figure~\ref{fig:AircraftDataNetwork}), and passengers’ (entertainment) network (figure~\ref{fig:Aircraft_Entertainment_Data_Network}).

Flight-control and/or crew networks consist of a multitude of sub-systems interconnected using wired technology as shown in figure~\ref{fig:AircraftDataNetwork}. These different avionics sub-systems are connected with end systems that are then interconnected by means of 
a backbone network using several communication standards like ARINC 429 \cite{ARINC429}, ARINC 825 \cite{ARINC825} or AFDX (ARINC 664 Aircraft Data Network, Part 7) \cite{ARINC664,li2009afdx}. Each arrow in the figure~\ref{fig:AircraftDataNetwork} represents a logical communication link that physically consists of two wires connecting these devices via different paths (dissimilar path redundancy). 

For some specific sub-systems, there are sets of sensors and actuators connected on Controller Area Network (CAN) \cite{Etschberger2001} or ARINC 429 buses for flight control systems \cite{Bartley2015}. The AFDX or a similar technology is used to interconnect time- and safety-critical sub-systems like environmental control, doors and other utility systems. The AFDX backbone also connects less critical sub-systems like displays providing safety information to passengers, but includes oxygen masks, oxygen flow and audio announcement triggers, and it manages the quality of service accordingly.

As shown in figure~\ref{fig:Aircraft_Entertainment_Data_Network}, the entertainment network can be supported using standard Ethernet technology -- it is not a high reliability- and safety-critical network.

The type and nature of the network configuration is dependent on the deployment scenario and objectives. However, there is a possibility that the flight control network, crew network and passenger (entertainment) network are all supported by the same wired technology, requiring implementation of network segregation by either physical separation of networks or by stringent firewalls, robust gateways and security policies \cite{Thanthry2004}.

\subsection{Benefits of Wireless Networks}
\label{sec:Benefits of Wireless Networks}
Whatever the network deployment topology and the communication technology used, one common element remains: the physical wire that connects two or more avionics sub-systems. Wiring an aircraft can be costly in that it includes wiring harness designs, cable fabrication and the associated exploitation cost due to the resulting additional weight. Furthermore, to provide dual redundancy these wires have to connect any two devices via two physically separated paths in the aircraft. Potentially, the wires and the related connectors represent 2-5 percent of an aircraft's weight \cite{ITU2010}. The design of the wiring route is heavily dependent on wiring harness design that has to satisfy the challenge of providing separate routing paths for redundant wiring. As the wiring of an aircraft is a time- and labor-intensive activity, post-deployment upgrades or installation of new wire routes or avionics sub-systems can be very expensive \cite{Dang2012}. As reported by \cite{ITU2010}, roughly 30 percent of wires are potential candidates for wireless substitutes. Therefore, as highlighted in \cite{RNAkram2015}, wireless solutions have reasonable prospects as long as security, safety and high levels of reliability can be maintained.

\section{Avionics Wireless Network}
\label{sec:Avionics_Wireless_Network}
\begin{figure}
	\centering
		\includegraphics[width=1.00\columnwidth]{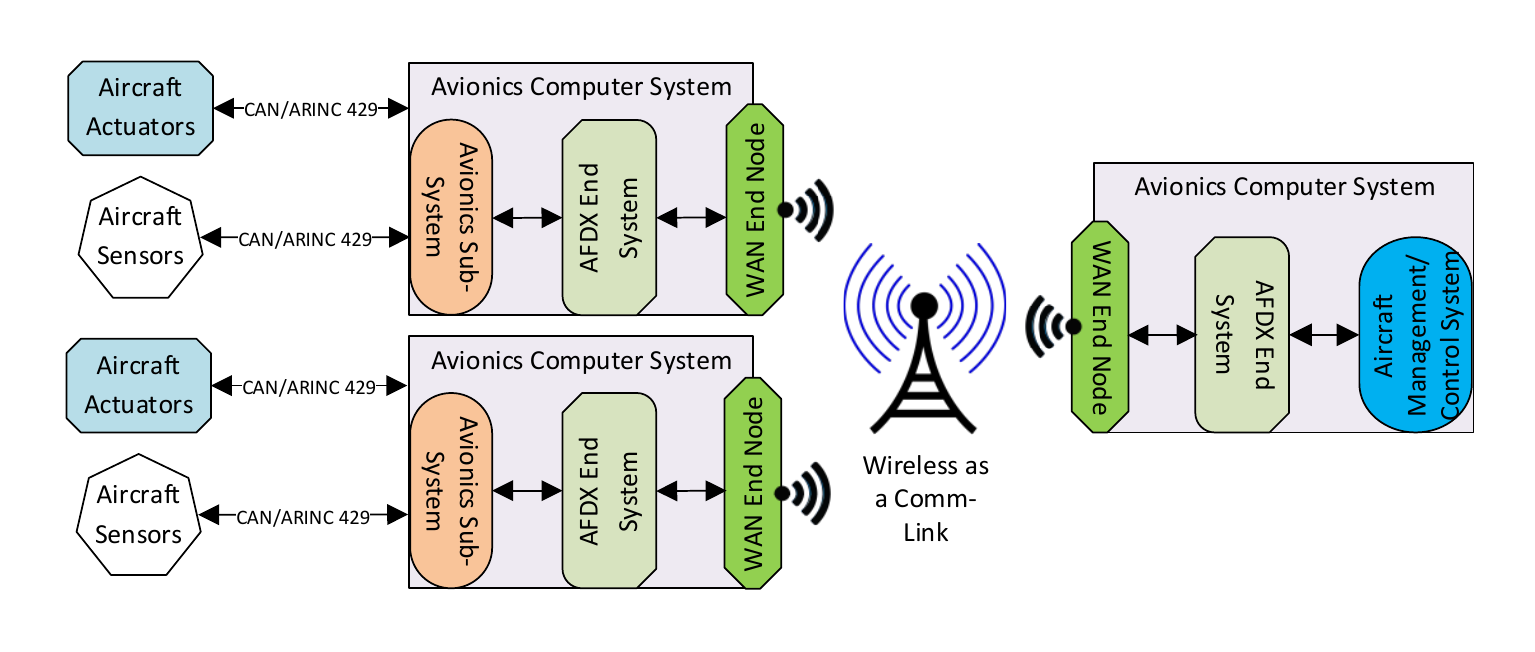}
	\caption{Generic Representation of Wireless as a Comm-Link}
	\label{fig:Wireless_as_a_Com-Link}
\end{figure}

Referring to \cite{RNAkram2015}, an AWN is defined as an aircraft network inter-connecting its different components using wireless technology instead of physical wires. Based on this definition, AWNs have been classified into four overlapping deployment architectures \cite{RNAkram2015}:

\begin{enumerate}[label=\textbf{\arabic*})]
\item {\bf Wireless Sensor Networks (WSN):} A WSN is a set of intelligent and autonomous systems that can sense physical or environmental conditions and/or act on them. Usually WSN nodes record the designated data and transfer them via a wireless medium to some dedicated nodes (so-called sinks) that act as gateways (as in a wireless mesh network) between the WSN and a third party system, to which the connection can be wired. As some related work has mentioned, such networks are particularly useful in aircraft design~\cite{Nickerson2001,Sampigethaya2009}, especially to monitor moving and/or rotating parts (for example, the landing gear~\cite{ITU2010} or the engine itself  ~\cite{Yedavalli2011}).  A WSN can also be useful as an independent network (\emph{i.e.} not connected to the aircraft network) simply to collect and store flight-related data; for instance, to improve the efficiency of an on-ground maintenance crew~\cite{Olive2006}.
\item {\bf Dissimilar Redundancy Network (DRN):} Since aircraft networks must be fault-tolerant, wireless links can be used to build dissimilar redundant networks; this would lower the probability of a potential common mode failure in networks based on the same wired technology. In addition, it decreases the difficulty of routing the wires - as much as possible, and as permitted by the aircraft geometry - using physically disjoint paths. Related work has identified that dissimilar network technologies can provide redundancy, which might enhance the overall reliability~\cite{Hitt2015} in some critical situations compared to ``identical redundancy"~\cite{downer2009failure}. 
\item {\bf Inflight Entertainment Network (IFE):} Since it is one of the least critical networks on board an aircraft, the Ethernet switch of the IFE depicted in figure~\ref{fig:Aircraft_Entertainment_Data_Network} can be replaced by a wireless access point as described by~\cite{Akl2010}, to offer more custom services without decreasing the overall safety of the aircraft.
\item {\bf Wireless as a Comm-Link:} In this type of AWN, wireless communication links replace the wired links between avionics computing modules as shown in Figure~\ref{fig:Wireless_as_a_Com-Link}. The protocols and the network architecture above the data link layer can remain the same, but at the physical layer, the data is communicated via a wireless medium rather than a wired medium. This type of network can be considered as a partial or a full deployment. In a partial deployment, out of two redundant wired links between aircraft sub-systems only one is replaced by a wireless link. In a full deployment both wired links between the aircraft sub-systems are replaced by wireless links. 
\end{enumerate}

\subsection{Related Work on Security Concerns}
\label{sec:Related Work on Security Concerns}
Security and trust have also been subject to analysis by both the academic community and the industry. A brief overview of aircraft information security and some improvements were proposed in \cite{Olive2006}. Security assurance research, from airplane production up to operation, was presented in \cite{Lintelman2006,ladstaetter2011security}. A general discussion on the security issues related to the aircraft network and aircrafts' internet connectivity is discussed in \cite{thanthry2006security}, while \cite{Thanthry2004,Robinson2007a} discuss the impact of WSNs deployed in aircraft and related security concerns. Security and safety are intrinsically related to each other in general and especially in the context of the aviation industry \cite{brostoff2001safe,Pfitzmann2004,Paulitsch2012}. The application and impact of cryptography, and especially the impact of public key cryptography for avionics networks, was evaluated in \cite{Robinson2007}.

Security and the general deployment of AWNs based on wireless-as-a-comm-link have been analyzed in \cite{RNAkram2015}, which discusses the security and trust challenges faced by AWNs. Beside this, a crucial component of the security of aircraft devices is the trusted boot process discussed in \cite{RNAkram2016a}. The security, trust and assurance issues related to bringing a user device into an aircraft network are evaluated in \cite{RNAkram2016b}.
 
\subsection{AWN Case Study}
\label{sec:Secure_High-Availability_Avionics_Wireless_Networks}
When considering the four deployment models discussed above for a wireless technology as part of an aircraft network, one thing is common: all of these options rely on data being transmitted over the air. This potentially makes it easier for an adversary to eavesdrop and/or modify the transmitted pieces of information. To prevent such eventualities, strong cryptography constructs are deployed to create a secure channel. A secure communication channel is encrypted with a key known only by the communicating entities. From an attacker's perspective he/she can still observe the encrypted messages but these should not give him/her any knowledge about the contents of the communication. Regarding the modification of messages, an attacker can modify them but the results (decrypted form of the transmitted message) would potentially not be in his/her control -- because he/she does not know the key used to encrypt the message. However, if a strong integrity mechanism is used to secure the channel, any modification would be detected.

Having secure channels for AWN communication is essential. To achieve this, an AWN has to run a secure channel protocol that would result in communicating entities being authenticated, and it also has to generate secure keys that would be used for encrypting the communications. For this reason, regardless of what type of AWN is chosen, it is necessary to set up a secure channel (to enable secure communication), which is the sole focus of this paper: our goal is to evaluate the security and performance of different secure channel protocols. We note that although wireless jamming and Denial-of-Service (DoS) attacks are valid and real concerns for AWNs we do not investigate them here and countermeasures to cope with such attacks are beyond the scope of this paper.

\section{Secure Channel Deployment Scenarios}
\label{sec:AWN_Deployment_Scenarios}
In this section, we discuss the three deployment scenarios that we have defined for the establishment of secure channels in AWNs. Even though these scenarios might not be exhaustive, we believe that they are representative. Wireless communication itself can be deployed either in Access Point (AP)  or ad-hoc modes. In this section, we are not concerned with AP and/or ad-hoc modes but with the nature of the key sharing mechanism, the supporting architecture, what is known prior to the execution of the protocol, and the execution of the secure channel protocols themselves. The issues of security and reliability directly related to the AP and the AP/ad-hoc modes are beyond the scope of this paper. 

\subsection{Pre-Shared Keys}
\label{sec:Pre-SharedKeys}
In this scenario, all communicating nodes in an AWN share a symmetric key that is provided either by their manufacturer or by the entity that deploys/configures them. This pre-shared key scenario can be achieved in two different ways. In the first method all the nodes have the same key to encrypt all the messages that they exchange with the other entities. In the second method, the pre-shared key is not used to directly encrypt the messages but to generate a session key. In this scenario the pre-shared key is referred to as the ``master key'' and a pre-defined algorithm is used to generate session keys.

\subsection{Trusted Key Distribution Frameworks (TKDF)}
\label{sec:TKDF}
In this scenario, all the nodes in the AWN trust a single entity that is responsible for generating and distributing the session keys that they will use to encrypt their messages. Such an entity is referred as a Trusted Key Distribution Server (TKDS) and all the nodes in the network have to communicate with it. Since not all nodes of the AWN might be in wireless communication range of the TKDS, they thus have to rely on neighboring nodes or on wireless range extenders (relay nodes). Each node in the AWN and the TKDS initially shares a secret key (either a symmetric or asymmetric key) and these keys are used to secure the communication between each of the nodes and the TKDS. 

\subsection{Key Sharing On Demand}
\label{sec:KeySharingOnDemand}
In this scenario, each individual node executes a secure channel protocol with its communicating peers in the network. The aim of this protocol is to authenticate the nodes with each other and to create session keys. For this deployment scenario, no prior sharing of keys is required as the session keys are computed during the execution of the protocol. For entity authentication, some prior knowledge of communicating partners is essential. This is required to successfully authenticate the entities. The process of authentication does not influence the key generation/sharing process in a protocol.

\section{Establishing Secure Wireless Connections}
\label{sec:Establishing_Secure_Channels}
The performance of several secure channel protocols has been measured over a wireless comm-link in a AWN test-bed. The results are presented in this section.

\subsection{Selected Protocols}
\label{sec:Selected Protocols}
 Seven different protocols have been considered to establish secure channels over the wireless comm-link of the AWN test-bed.
They can be gathered in three different families: secure channels based on pre-shared keys; secure channels based on Trusted Key Distribution Frameworks (TKDFs); and secure channels based on key sharing protocols, as discussed in section \ref{sec:AWN_Deployment_Scenarios}.

For secure channels based on pre-shared keys, Wired Equivalent Privacy (WEP), Wi-Fi Protected Access Pre-Shared Key (WPA-PSK) and IPSec have been selected.
When using WEP, each node has a fixed pre-shared key used to encrypt the data frames using RC4.
With WPA-PSK, each node has a pre-shared master key that is used to build session keys during the authentication phase.
Compared to the two previous secure channel protocols, IPSec encryption (with fixed pre-shared keys in our experiments) is achieved at the level of layer 3 (\emph{i.e.} network) of the OSI protocol stack instead of layer 2 (\emph{i.e.} data link) for WEP and WPA-PSK.

For secure channels based on TKDFs, two ad-hoc frameworks were developed. The authentication and key distribution phase is based on symmetric keys for the first framework whereas it is based on asymmetric keys for the second framework.
In a symmetric key-based TKDF, each node shares a symmetric key with a trusted key distribution server, which uses it to send the session key encrypted with the shared key associated with each communicating node. Then each node  decrypts it to use the key to finalize the establishment of the secure channel. For our experiments, the fixed version of the Needham-–Schroeder Symmetric Key Protocol, which is the basis of Kerberos, was implemented using AES as the symmetric encryption algorithm.

In an asymmetric key-based TKDF, each node shares a public key with a trusted key distribution server, which uses it to send encrypted session keys that the target node deciphers with its private key. For our experiments, the fixed version of the Needham-–Schroeder (so-called Needham-–Schroeder--Lowe) public Key Protocol was implemented using RSA as the public key algorithm. 
In both TKDFs, the distributed session keys are symmetric keys used by the parties to communicate in the network. In our implementations, AES was used as the channel encryption algorithm.

For the secure channels based on key sharing protocols, SSH and SSL were selected.

\subsection{Comparison Criteria}
\label{sec:Comparison_Criteria}

For a protocol to support the AWN framework, it should meet, at minimum, the security and operational requirements listed below:

\begin{enumerate}[label=\textbf{G\arabic*})]

\item {\bf {\itshape Mutual Entity Authentication}:} All nodes in the network should be able to authenticate with each other so as to avoid masquerading by a malicious entity.

\item {\bf Asymmetric Architecture:} Certified public keys should be exchanged between the entities to facilitate the key generation and entity authentication process. 

\item {\bf Mutual Key Agreement:} Communicating parties should agree on the generation of a key during the execution of the protocol. 

\item {\bf Joint {\itshape Key Control}:} Communicating parties should mutually control the generation of new keys to prevent one party from choosing weak keys or predetermining any portion of the session key.  

\item {\bf {\itshape Key Freshness}:} The generated key should be fresh with regards to the protocol session to prevent replay attacks.

\item {\bf Mutual {\itshape Key Confirmation}:} The communicating parties should provide implicit or explicit confirmation that they have generated the same keys during a run of the protocol. 

\item {\bf Known-Key Security:} Should a malicious user obtain the session key of a particular protocol run, he/she should not be able to retrieve long-term secrets ({\itshape private keys}) or {\itshape session keys} (future and past).

\item {\bf Unknown {\itshape Key} Share Resilience:} In the event of an unknown key share attack, an entity $\mathcal{X}$ believes that it has shared a key with $\mathcal{Y}$, where the entity $\mathcal{Y}$ mistakenly believes that it has shared the key with entity $\mathcal{Z} \neq \mathcal{X}$. The proposed protocols should adequately protect against this kind of attack.

\item {\bf {\itshape Key} Compromise Impersonation (KCI) Resilience:} If a malicious user retrieves the long-term key of an entity $\mathcal{Y}$, it will enable him to impersonate $\mathcal{Y}$. Nevertheless, compromising the key should not enable him to impersonate other entities\cite{Blake-Wilson:1997:KAP:647993.742138}.

\item {\bf {\itshape Perfect Forward Secrecy}:} If the long-term keys of communicating entities are compromised, this should not enable a malicious user to compromise previously generated session keys. 

\item {\bf Mutual {\itshape Non-Repudiation}:} The communicating entities will not be able to deny that they have executed a protocol run with each other. 

\item {\bf Partial Chosen Key (PCK) Attack Resilience:} Protocols that claim to provide joint key control are susceptible to this type of attack \cite{Mitchell1998}. In this type of attack, if two entities provide separate values to the key generation function then one entity has to communicate its contribution value to the other. The second entity can then compute the value of its contribution in such a way that it can dictate its strength (i.e.\ it is able to generate a partially weak key). This attack depends upon the computational capabilities of the second entity. The proposed protocols should adequately prevent PCK attack.


\item {\bf Privacy:} A third party should not be able to know the identities of the AWN nodes.

\end{enumerate}

For a formal definition of the (italicized) terms used in the above list, the reader is referred to \cite{Menezes1996}. The requirements listed above are used below as a point of reference to compare the selected protocols in Table \ref{tab:ProtocolComparisonOnTheBasiesOfStatedGoals}.

For the performance evaluation that we have conducted, the main measurements are related to the time required to establish a secure channel once the wireless link is established (or to establish the secure wireless link for protocols like WEP and WPA-PSK operating at the level of the data link layer).
The properties of the keys (e.g. type of keys, key size, and freshness) will be discussed with regards to the performance results.

\begin{figure}[htbp]
	\centering
		\includegraphics[width=1.00\columnwidth]{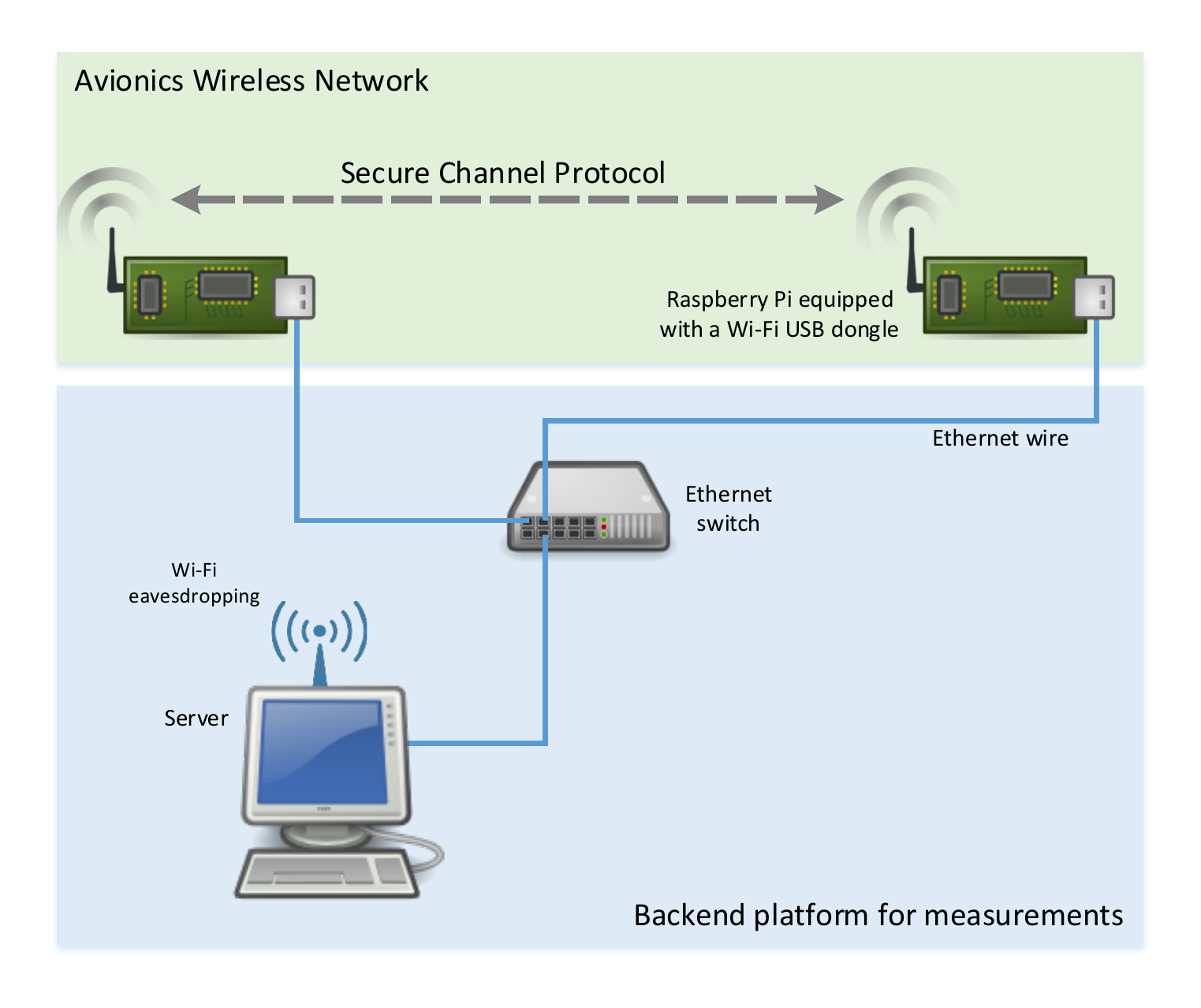}
	\caption{AWN test-bed}
	\label{fig:AWN test-bed}
\end{figure}

\subsection{Test-Bed for Performance Evaluation}
\label{sec:Test-Bed}
In our AWN test-bed each node is a Raspberry Pi model B supplied with a Wi-Fi USB dongle TL-WN722N by TP-LINK.
In all our measurements, the nodes were configured in ad-hoc mode.

For all the selected protocols, in our evaluation implementations, only 2 nodes establish a secure channel.
However, for the TKDF, a key distribution server is also required and a third node in the ad-hoc network plays this role.

In our AWN test-bed, each node is connected to a backend server by means of an Ethernet connection.
This server controls the nodes so as to prepare them for the target scenario and is also in charge of collecting the measurements.
Effective measurement can be done internally on the node initiating the secure channel, called a client, or at the level of the network data exchanged between the nodes of the AWN and captured with a Wi-Fi card on the backend server set in monitor mode.

\section{Security and Performance Analysis}
\label{sec:Security_and_Performance_Analysis}
In this section, we present the security analysis of the selected protocols based on the goals stated above; it is followed by the performance analysis of these protocols. We conclude the section with an overall analysis and with a discussion of some future research directions. 

\subsection{Security Analysis}
\label{sec:Security_Analaysis}
Before discussing the analysis of the security goals as met by each protocol presented in Table~\ref{tab:ProtocolComparisonOnTheBasiesOfStatedGoals}, it is worth noting that several of them can be configured in several manners that may change the way they satisfy the goals and may also change their performance. For instance, we decided to use IPSec with fixed pre-shared keys. This choice is not one that can satisfy the maximum number of goals but this solution is more suitable for resource-constrained wireless nodes. However, Internet Key Exchange (IKE) protocol could have been used and then IPSec would have met additional goals. 

When comparing the selected protocols, taking into account the above remark, it is interesting to note that the protocols based on asymmetric cryptosystems are those that meet most of the goals. However, these solutions are known to be costly in terms of time and resource consumption, as confirmed by the performance measurements presented in the following section.
The secure channel protocols acting in the low level layers of the OSI stack, like WEP, WPA-PSK (layer 2), or even IPSec (layer 3), fail to satisfy several goals, which is surprising because, usually, security solutions provided at low levels are more generic; they might be expected to ensure better security than solutions working at higher levels. Solutions at these levels that establish more secure channels do exist (e.g. 802.1X and RADIUS server at level 2, IPSec and PKI) but they are not applicable to resource-constrained wireless nodes.
Thus, among our selected protocols, solutions that establish secure channels at higher levels (SSH or SSL operate respectively at levels 7 and 5-7) and/or that rely on a server (symmetric and asymmetric TKDF) satisfy more goals. However these solutions are too costly: either they are based on asymmetric cryptosystems (which are costly in terms of resources) or they need a server that is costly in terms of bandwidth, latency and delay.

\begin{table}[h]
\caption{Protocols comparison on the basis of the stated goals (see section \ref{sec:Comparison_Criteria}.)} 
				\label{tab:ProtocolComparisonOnTheBasiesOfStatedGoals}

\begin{center}
\resizebox{0.95\columnwidth}{!}{%
			\begin{tabular}{!{\vrule width 1pt}@{ }r@{ }l!{\vrule width 0.75pt}c|c|c|c|c|c|c!{\vrule width 1pt}}
\noalign{\hrule height 1pt}
 			\multicolumn{2}{!{\vrule width 1pt}l!{\vrule width 0.75pt}}{\multirow{2}{*}{\bf Goals}} &  \multicolumn{7}{c!{\vrule width 1pt}}{\bf Protocols }  \\ \cline{3-9}
	    && WEP &   WPA-PSK  & IPSec &  Symmetric TKDF & Asymmetric TKDF &  SSH & SSL   \\ 
		 \noalign{\hrule height 0.75pt}
 		G1. && $-*$  & $-*$ & $-*$  & $*$   & $*$   & ($*$) & $*$ \\ \hline
        G2. && $\x$  & $\x$	& $\x$  & $\x$  & $*$   & $*$   & $*$   \\ \hline
        G3. && $\x$  & $-*$ & $\x$  & $-*$  & $-*$  & $*$   & $*$   \\ \hline
        G4. && $\x$  & $-*$ & $-*$  & $-*$  & $-*$  & ($*$) & ($*$) \\ \hline
        G5. && $\x$  & $*$  & $\x$  & $*$   & $*$   & $*$   & $*$   \\ \hline
        G6. && $*$   & $*$  & $*$   & $*$   & $*$   & $*$   & $*$   \\ \hline
        G7. && $\x$  & $*$  & $\x$  & $*$   & $*$   & $*$   & $*$   \\ \hline
        G8. && $-*$  & $-*$ & $-*$  & $-*$  & $-*$  & $*$   & $*$   \\ \hline
        G9. && $\x$  & $\x$ & $*$   & $*$   & $*$   & $*$   & $*$   \\ \hline
        G10.&& $\x$  & $\x$ & $\x$  & $*$   & $*$   & $*$   & $*$   \\ \hline
        G11.&& $\x$  & $\x$ & $\x$  & $\x$  & $*$   & $*$   & $*$   \\ \hline
        G12.&& $*$   & $*$  & $*$   & $*$   & $*$   & $*$   & $*$   \\ \hline
        G13.&& $-*$  & $-*$ & $-*$  & ($*$) & ($*$) & ($*$) & ($*$)  
        \\ \noalign{\hrule height 0.75pt}
		\end{tabular}
        }
\end{center}		
	{\bf Note: }{$*$ means that the protocol meets the stated goal, ($*$) shows that the protocol meets the requirement in certain conditions,  $\x$ shows that the protocol cannot meet the stated goal and $-*$ means that the protocol (implicitly) meets the requirement, not because of the protocol messages but because of the prior relationship between the communicating entities.}

\end{table}

\subsection{Performance Analysis}
\label{sec:Performance_Analysis}
The practical results obtained on our AWN test-bed confirm the theoretical analysis: asymmetric cryptosystems are costly and solutions relying on a third party (TKDFs) are even more costly.
To be fair, it is important to note that we implemented the two TKDF protocols ourselves whereas the implementations of other tested protocols were done by groups of professional developers.
Thus our implementations may be optimized, but not to the point where this would change the results by a significant factor.
In addition, it can be noted that SSL and SSH operate over a TCP connection, which is more time consuming for the establishment of communication than our implementations of TKDF, which operate over UDP to improve performance (at the expense of the reliability of the connection).

The good performance of WEP and WPA-PSK are related to the fact that the protocols are run by the dedicated hardware and firmware of the Wi-Fi card - optimized for the execution of these protocols. The good results for IPSec are also related to the use of optimized hardware on the Wi-Fi card to execute this protocol - as it is an important protocol for Internet communication.

\begin{table}[h]
\caption{Performance comparison of selected secure channel protocols.}
\label{tab:Performance comparison of selected secure channel protocols.}

\begin{center}
\resizebox{0.95\columnwidth}{!}{%
\begin{tabular}{|c|c|c|c|}
\hline
\bf Protocol & \bf Key type &\bf Key size (bits)& \bf Establishment (ms)\\ \hline
WEP & RC4 & 128 &2.42\\ \hline
WPA & AES & 128 & 2.55\\ \hline
IPSec & AES & 256 & 2.67\\ \hline
Symmetric TKDF & AES & 256 & 5092.88\\ \hline
Asymmetric TKDF & RSA & 2048 & 14447.63 \\ \hline
SSH & RSA & 2048 & 911.21\\ \hline
SSL & RSA & 2048 & 1310.93 \\ \hline
\end{tabular}
}
\end{center}
\end{table}

Note that in WEP and WPA-PSK in ad-hoc mode, the packet loss was very important, respectively around 50\% and 70\%.
WPA-PSK in AP mode needs the same time to establish the secure channel but the rate of packet lost was only 20\%.
For IPSec in ad-hoc mode, which encrypts at layer 3 over a plain text channel at layer 2, the rate of packet loss was 0\%.
Thus, as mentioned in the next section, the connection mode and the layer at which the secure channel should be established are parameters that should be studied in future work.

\subsection{Overall Analysis and Future Research Directions}
\label{sec:Overall_Analysis_and_Future_Research_Directions}

It appears that the selected protocols (all state-of-the-art in computer science security) are too generic, \emph{i.e.} not specifically tailored for the target applications, or do not offer acceptable performance. Therefore, as part of our future research, we are currently experimenting with:

\begin{itemize}
\item A new secure and trusted channel protocol that meets all the stated requirements, moving away from large API (Application Programming Interface) based protocols like SSL and IPSec that might introduce implementation-related vulnerabilities.
\item Security and reliability of AP and ad-hoc modes in different AWN deployment contexts.
\end{itemize}

Additionally, we are exploring the following directions:

\begin{itemize}
\item Countering wireless jamming and DoS attacks.
\item Secure execution on nodes using ARM TrustZone and Intel SGX.
\end{itemize}

\section{Conclusion}
\label{sec:Conclusion}

In this paper, we have discussed the nature of ADN and how AWNs might provide a valid alternative to wired networks. Any communication that uses a wireless medium has the inherent issue that an attacker can easily access this physical communication link. This can enable the attacker to eavesdrop and/or modify the contents of messages. To avoid this, secure channels are essential to encrypt all messages. For this encryption to be secure and robust, the keys that are used need to be not only secure but also to meet additional security requirements. In this paper, we listed thirteen security goals that we believe any secure channel protocol should meet. Subsequently, we selected seven different secure channel protocols (representative examples from three different wireless deployment scenarios). We developed a test-bed to evaluate their performance. We then compared the seven selected protocols in terms of security and performance.

There is no doubt that extensive work is still required before an AWN can be deployed in an aircraft environment and there are many challenges to overcome. However, in this paper we provide security comparisons and experimental performance data that we believe will be useful for someone wanting to deploy an AWN  to enable them to make an informed decision about which features/protocols meet their unique environment and requirements. This paper contributes to the work that needs to be done to make AWNs a robust and secure proposal.

\section*{Acknowledgments}

The authors from Royal Holloway University of London acknowledge the support of the UK's innovation agency, InnovateUK, and the contributions of the SHAWN project partners. The authors from XLIM acknowledge the support of:
\begin{itemize}
\item the SFD (Security of Fleets of Drones) project funded by R\'egion Limousin;
\item the TRUSTED (TRUSted TEstbed for Drones) project funded by the CNRS INS2I institute through the call 2016 PEPS (``Projet Exploratoire Premier Soutien'') SISC (``S\'ecurit\'e Informatique et des Syst\`emes Cyberphysiques'');
\item the SUITED (Suited secUrIty TEstbed for Drones) and UNITED (United NetworkIng TEstbed for Drones) projects funded by the MIRES (Math\'ematiques et leurs Interactions, Images et information num\'erique, R\'eseaux et S\'ecurit\'e) CNRS research federation;
\end{itemize}
The authors from LaBRI acknowledge the support of:
\begin{itemize}
\item the TRUSTED (TRUSted TEstbed for Drones) project funded by the CNRS INS2I institute through the call 2016 PEPS (``Projet Exploratoire Premier Soutien'') SISC (``S\'ecurit\'e Informatique et des Syst\`emes Cyberphysiques'');
\item the SUITED-BX and UNITED-BX projects funded by LaBRI and its MUSe team.
\end{itemize}

\section*{Disclaimer}
The views and opinions expressed in this article are those of the authors and do not necessarily reflect the position of SHAWN project or any of organisations associated with this project. 

\bibliographystyle{IEEEtran}
\bibliography{main}



\end{document}